%% file: internal_heating_rapids.tex
\documentclass[paper]{JFM-FLM_Au}

\usepackage{graphicx,footmisc}
\usepackage[labelformat=simple]{subcaption}

\NewDocumentCommand{\subfloat}{o o m}{\subcaptionbox{}{#3}\ignorespaces}
\usepackage{epstopdf}
\usepackage{amsmath}
\usepackage{color}
\usepackage{placeins}

\def\be{\begin{equation}}
\def\ee{\end{equation}}

\def\lb{\left ( }
\def\rb{\right ) }
\def\ep{\epsilon}

\def\dst{{\partial_t}}
\def\dZ{{\partial_Z}}

\def\Rat{\widetilde{Ra}}

\def\Ret{\widetilde{Re}}
\newcommand{\dnsfig}{\ref{F:dns}}

\righttitle{Internal heating in rapidly rotating convection}
\lefttitle{J. A. Nicoski, T. G. Oliver, S. Maffei and M. A. Calkins}

\title{Internal heating in rapidly rotating convection is not a shortcut to geostrophic turbulence}

\author[J. A. Nicoski, T. G. Oliver, S. Maffei and M. A. Calkins]
{Justin A. Nicoski\aff{1},
 Tobias G. Oliver\aff{1},
 Stefano Maffei\aff{2}
 and Michael A. Calkins\aff{1}}

\affiliation{\aff{1}Department of Physics, University of Colorado, Boulder,
Colorado 80309, USA
\aff{2}Institute of Geophysics, ETH Zurich, Sonneggstrasse 5, 8092 Zurich,
Switzerland}

\corresau{Michael A. Calkins, \email{michael.calkins@colorado.edu}}

\begin{document}

\maketitle

\begin{abstract}
Convective turbulence in planets and stars is often driven by internal heating. This forcing mechanism has also been proposed as a means of accessing the diffusion-free scalings of the geostrophic turbulence (GT) regime at modest forcing. We test this with the asymptotically reduced quasi-geostrophic model, which is formally valid in the limit of vanishing Ekman number, $Ek \rightarrow 0$, and retains the Prandtl number as an independent parameter. We find no such shortcut: the Nusselt and Reynolds numbers are no closer to their diffusion-free predictions than in the boundary-heated case; the predicted $Pr^{-1/2}$ collapse fails over the range surveyed; and the prefactor depends on the form of the heating. Our no-slip/stress-free cases track the radiatively driven simulations of Hadjerci, Bouillaut, Miquel \& Gallet (\textit{J. Fluid Mech.}, vol. 998, 2024, A9) case by case. Those simulations span only the narrow range of forcing over which the compensated Nusselt number is near its maximum within their geostrophic-turbulence criterion; across the wider range accessible to the reduced model it rises to that maximum and then falls by more than an order of magnitude while the flow remains rotationally constrained. We argue that bulk transport scalings are incomplete diagnostics of GT, whereas the saturation of the interior mean temperature gradient and of the vertical velocity kurtosis remain reliable indicators.
\end{abstract}


\section{Introduction}
\label{S:intro}

Buoyancy-driven turbulence is a fundamental heat transport process in the rotating fluid layers of stars and planets \citep{jmA15}. The canonical model with which to study these flows is rotating Rayleigh--B\'enard convection (RRBC), in which a rotating fluid layer is heated from below and cooled from above \citep{rE23}. Most prior work imposes constant-temperature or constant-heat-flux conditions at the boundaries, whereas convection in natural systems is more commonly driven by internal heat sources -- radiogenic decay and secular cooling in planetary mantles and cores, absorbed radiation in atmospheres -- for which neither boundary-driven configuration is a faithful proxy.

The regime of interest is geostrophic turbulence (GT), in which the leading-order force balance is geostrophic but the bulk is nevertheless fully turbulent \citep{rK21}. \cite{kJ12} identified three behaviours associated with GT: the breakdown of coherent convective plumes, the saturation of the kurtosis of the vertical velocity, and the saturation of the mean temperature gradient to a non-zero value. In the GT regime the transport is expected to follow the diffusion-free scalings
\be
Nu \sim \Rat\vphantom{Ra}^{3/2} Pr^{-1/2}, \qquad
\Ret \sim \Rat Pr^{-1},
\label{e:df}
\ee
where $\Rat = Ra\,Ek^{4/3}$ is the reduced Rayleigh number and $\Ret = Re\,Ek^{1/3}$ the reduced Reynolds number \citep{kJ12b,sM21}. Access to these scaling regimes appears easiest at small Prandtl number \citep{kJ12b,jA26}. These same transport scalings follow from the rotating mixing-length theory of \cite{aB14} and \cite{lC20}, though their temperature-gradient scaling incorrectly predicts an isothermal interior at large Rayleigh number \citep[e.g.][]{bM26}. Here $Ra = g\alpha\Delta T H^{3}/\nu\kappa$ and $Ek = \nu/2\Omega H^{2}$ are the Rayleigh and Ekman numbers for a layer of depth $H$ rotating at rate $\Omega$, where $g$ is the gravitational acceleration, $\alpha$ the thermal expansion coefficient, $\Delta T$ the time- and horizontally averaged temperature difference across the layer, and $\nu$ and $\kappa$ the kinematic viscosity and thermal diffusivity; $Pr = \nu/\kappa$ is the Prandtl number. The Nusselt number $Nu$ is the ratio of the total to the conductive heat flux, and $Re$ is the Reynolds number based on the layer depth and a characteristic flow speed. ``Diffusion-free'' and ``GT'' are not interchangeable: GT demands rotational constraint, whereas diffusion-free scaling is a statement about the independence of the dimensional heat flux and flow speed from $\nu$ and $\kappa$. Neither regime is uniquely identified by the exponents alone, which ignore the $Pr$ dependence and may hold over limited ranges of parameter space in which viscosity nevertheless influences the flow \citep{sM21}. For this reason \cite{jS24,jS24b} require the diffusion-free scalings of the heat transport, the flow speed and the convective length scale to hold simultaneously before a flow is identified as GT, and \cite{vK25b} find that with stress-free boundaries the heat transport scaling is attained well before the other two.

It has been suggested that diffusion-free flows are more easily accessed with internal heating than with boundary heating. In non-rotating convection the diffusion-free state is expected to coincide with the turbulent breakdown of the boundary layers \citep{rK62}; internal heating removes the thermal boundary layers, and numerical \citep{sL18,sK22} and experimental \citep{vB19} results confirm that the diffusion-free regime is then reached more readily. Whether the same holds under rotation is less clear \textit{a priori}. Prior studies have shown that the flow undergoes a sequence of regime transitions as the Rayleigh number is increased, from cellular convection through convective Taylor columns and plumes to geostrophic turbulence \citep{kJ12,dN14}. Each transition is controlled by $\Rat$ and $Pr$ and occurs in the interior. Thermal boundary layers form as in non-rotating convection; however, \cite{kJ96,kJ12b} found that the presence of an interior mean temperature gradient limits the extent to which the thermal boundary layers control the heat transport. In fact, the influence that the thermal boundary layers exert on the heat transport is a \textit{decreasing} function of Rayleigh number. Thus, removing the influence of the thermal boundary layers via the introduction of internal heating may not be sufficient to allow for a more rapid transition to the GT regime.

In rotating experiments, centrifugal buoyancy restricts laboratory cells to $Ek \gtrsim 10^{-7}$ \citep{sH18,sH19}, and at these Ekman numbers boundary-heated RRBC has reached a diffusion-free regime only at the small Prandtl numbers of liquid metals \citep{jA26}. Nevertheless, the radiatively driven experiments and simulations of \cite{vB21} and \cite{gH24} report diffusion-free behaviour in all bulk quantities at modest forcing, even at $Pr = 7$, where boundary-heated RRBC does not reach GT in the accessible parameter range. More recent direct numerical simulations (DNS) in 2.5D Cartesian \citep{tJ25} and spherical-shell \citep{nL26} geometries report a more nuanced picture, with diffusion-free heat transport but diffusion-dependent velocity and zonal-flow scalings.

Whether internal heating accesses diffusion-free GT scalings more readily is, by construction, a question about the limit $Ek \rightarrow 0$. By a ``shortcut'' we mean that the GT regime, and the diffusion-free scalings that characterise it, would be reached at a smaller reduced Rayleigh number $\Rat$ than would be the case in a boundary-heated system. We therefore use the asymptotically reduced quasi-geostrophic (QG) model of rapidly rotating convection \citep{kJ98a,mS06,kJ16}, derived as the leading-order balance in this limit with $\ep = Ek^{1/3} = Ro$ as the expansion parameter. The model is defined on the $Ek \rightarrow 0$ manifold, retains $Pr$ as an explicit independent parameter, and admits values of $\Rat$ well beyond what DNS at finite $Ek$ can resolve \citep{tO23}. Recent rescaled-equation DNS at small Rossby number have confirmed quantitative agreement with the asymptotic model \citep{aK25,kJ25}. The diffusion-free prediction is therefore a directly testable property of the model, and we use it to show that internal heating provides no shortcut to it.

\section{Model}
\label{S:model}

We consider rotating convection in a horizontally periodic Cartesian layer of depth $H$, characterised by the fast coordinates $(x,y,z)$ and the slow axial coordinate $Z = \ep z$. Internal heating is rigorously captured in the QG model by requiring that the heating profile vary only over $Z$. The non-dimensional equations \citep{kJ98a,kJ16,mP16} are
\begin{subequations}
	\begin{equation}
		\dst w + \mathcal{J}(\psi,w) + \partial_Z \psi
		= \frac{\Rat_q}{Pr} \theta + \nabla_\perp^2 w,
		\label{e:mw}
	\end{equation}
	\begin{equation}
		\dst \zeta + \mathcal{J}(\psi,\zeta) - \partial_Z w
		= \nabla_\perp^2 \zeta,
		\label{e:mvz}
	\end{equation}
	\begin{equation}
		\dst \theta + \mathcal{J}(\psi,\theta) + w\partial_Z \overline{T}
		+ \ep\!\left[\nabla_\perp\!\cdot\!\left(\mathbf{u}_{1\perp}^{ag}\theta\right) + \partial_Z\!\left(w\theta - \overline{w\theta}\right)\right]
		= \frac{1}{Pr}\!\left(\nabla_\perp^2 + \ep^2 \partial_Z^2\right)\theta,
		\label{e:ft}
	\end{equation}
	\begin{equation}
		\partial_Z\left( \overline{w\theta}\right)
		= \frac{1}{Pr}\partial_Z^2 \overline{T} + \frac{1}{Pr} Q(Z),
		\label{e:meanheat}
	\end{equation}
	\label{e:qg}
\end{subequations}
together with $\mathbf{u}_{1\perp}^{ag} = -\partial_Z \nabla_\perp \phi$ and $\nabla_\perp^2 \phi = w$. Here $w$ is the vertical velocity, $\zeta$ the vertical vorticity, $\psi$ the geostrophic streamfunction, $\theta$ the fluctuating temperature, $\overline{T}$ the horizontally averaged temperature, $Q(Z)$ the heating profile and $\mathcal{J}(f,g) = \partial_x f\,\partial_y g - \partial_x g\,\partial_y f$. Equations \eqref{e:qg} constitute the composite model of \cite{kJ16}: the $O(\ep)$ advective and $O(\ep^2)$ diffusive terms in \eqref{e:ft} are formally higher order in the fluid interior, but Ekman pumping at a no-slip boundary drives thermal fluctuations that become large within an $O(Ek^{1/3}H)$ thermal wind layer, where the nominally higher-order terms are promoted to leading order; retaining them regularises this boundary-layer response and captures the pumping-enhanced heat transport. 
The control parameters are the Prandtl number and the reduced flux-based Rayleigh number,
\be
\Rat_q = \ep^4 Ra_q = \ep^4\frac{g\alpha q H^4}{\nu \kappa},
\ee
where $q$ is the prescribed heating flux. The heat flux through the layer is therefore imposed in every case considered here. The temperature difference $\Delta T$ between the vertical boundaries is an output of the simulation. We consider $Pr = 1$ and $7$. Note that $Ra_q$ corresponds to the flux-based Rayleigh number $Ra_P$ of \cite{vB21} and \cite{gH24}.

The boundaries are impermeable and we consider two configurations: doubly stress-free (SF-SF), for which we set $\ep = 0$ and recover the model of \cite{kJ98a}; and no-slip lower/stress-free upper (NS-SF), which mimics the radiatively driven experiments of \cite{vB21} and \cite{gH24} and incorporates the no-slip condition through parametrised Ekman pumping, $w = (\ep^{1/2}/\sqrt{2})\,\zeta$ at $Z=0$. The fluctuating temperature satisfies $\theta = 0$ at both boundaries and the mean temperature has zero flux at $Z=0$; the no-secular-heating condition then fixes $\partial_Z \overline{T} = -F(1)$ at $Z=1$, where $F(Z) = \int_0^Z Q(\xi)\,d\xi$ is the cumulative heating flux. The condition $\theta = 0$ is compatible with insulating boundaries: \cite{mC15c} showed that in the rapidly rotating limit fixed-flux and fixed-temperature thermal boundary conditions are asymptotically equivalent. The flux condition is carried entirely by the mean temperature, while the fluctuating temperature satisfies $\theta = 0$ at leading order.

We use two heating profiles. The first is motivated by the radiative absorption profile of \cite{vB21} and \cite{gH24} and is built from the normalised exponential $\mathcal{E}(\delta,Z) = e^{-Z/\delta}/[\delta(1-e^{-1/\delta})]$. We refer to $Q(Z) = \mathcal{E}_c(0.05,Z) = \mathcal{E}(0.05,Z) - 1$ as the ``Exponential'' profile; the second, constant term is a uniform cooling term that prevents secular heating and makes both boundaries insulating. We also consider a ``Uniform'' profile, $Q(Z) = 1$, with a fixed cooling flux at $Z=1$. A linear stability analysis of \eqref{e:qg}, solved via a sine expansion in $Z$, gives critical values of the reduced Rayleigh number and horizontal wavenumber of $(\Rat_{qc}, k_c) = (16.61, 1.321)$ for the Exponential profile and $(16.39, 1.323)$ for the Uniform profile, against $(8.6956, 1.3048)$ for the boundary-heated case.

The Nusselt number is the ratio of total to conductive heat flux \citep{dG16b}, $Nu = \Delta T^{-1}\int_0^1 F(Z)\,dZ$, where $\Delta T$ is the time- and horizontally averaged temperature difference between the boundaries; the Uniform and Exponential profiles give $Nu = 1/(2\Delta T)$ and $Nu = [(1-e^{-1/\delta})^{-1} - \delta - 1/2]/\Delta T$ respectively. We report results in terms of the fixed-temperature reduced Rayleigh number $\Rat = \Delta T\,\Rat_q$, enabling direct comparison with boundary-heated studies. We also report the vertical Reynolds number $\widetilde{Re}_z = \langle w^2 \rangle^{1/2}$ and the total Reynolds number $\widetilde{Re} = \langle \mathbf{u}\cdot\mathbf{u} \rangle^{1/2}$, formed from all three velocity components, where the angle brackets denote a volume and time average. The equations are solved with a Cartesian version of the QuICC code, the spherical implementation of which is described by \cite{pM16}. The spatial discretisation is pseudospectral, with Fourier series in the two periodic horizontal directions and Chebyshev polynomials in $Z$; the vertical operators are inverted with the sparse quasi-inverse method of \cite{kJ09}. Time integration uses the third-order implicit--explicit Runge--Kutta scheme of \cite{pS91}, in which the linear terms are treated implicitly and the nonlinear terms explicitly. The horizontal extent is $L_x = L_y = 10\lambda_c$, where $\lambda_c = 2\pi/k_c$. The SF-SF cases span $12 \le \Rat \le 260$ on the $Ek = 0$ manifold. The NS-SF cases, all at $Pr = 7$, comprise a scan at fixed $Ek = 10^{-7}$ over $10 \le \Rat \le 236$ and four cases at $Ek = 5\times10^{-7}$, $7\times10^{-7}$, $10^{-6}$ and $1.5\times10^{-6}$ with $Ra_q = 10^{12}$, matched to the DNS series of \cite{gH24}. The full set of simulations, with resolutions and time steps, is tabulated in Appendix~\ref{A:table}.

The composite model has been validated against no-slip DNS at $Ek$ down to $10^{-7}$ \citep{kJ16,mP16}, and \S\ref{S:comparison} adds a validation against the radiatively driven DNS of \cite{gH24} at $Ek$ between $5\times10^{-7}$ and $1.5\times10^{-6}$. Its asymptotic validity requires the local Rossby number $\ep\,\Ret$ to remain small. Along our $Ek = 10^{-7}$ scan it reaches $0.18$ at the largest $\Rat$, and in this regime \cite{aK25} find that DNS at $Ek = 10^{-7}$ are not fully converged to the rotationally constrained state.

\section{Results}
\label{S:res}

\subsection{Transport scalings}
\label{S:scalings}

\begin{figure}
\begin{center}
\subfloat[][]{\includegraphics[width=0.5\textwidth]{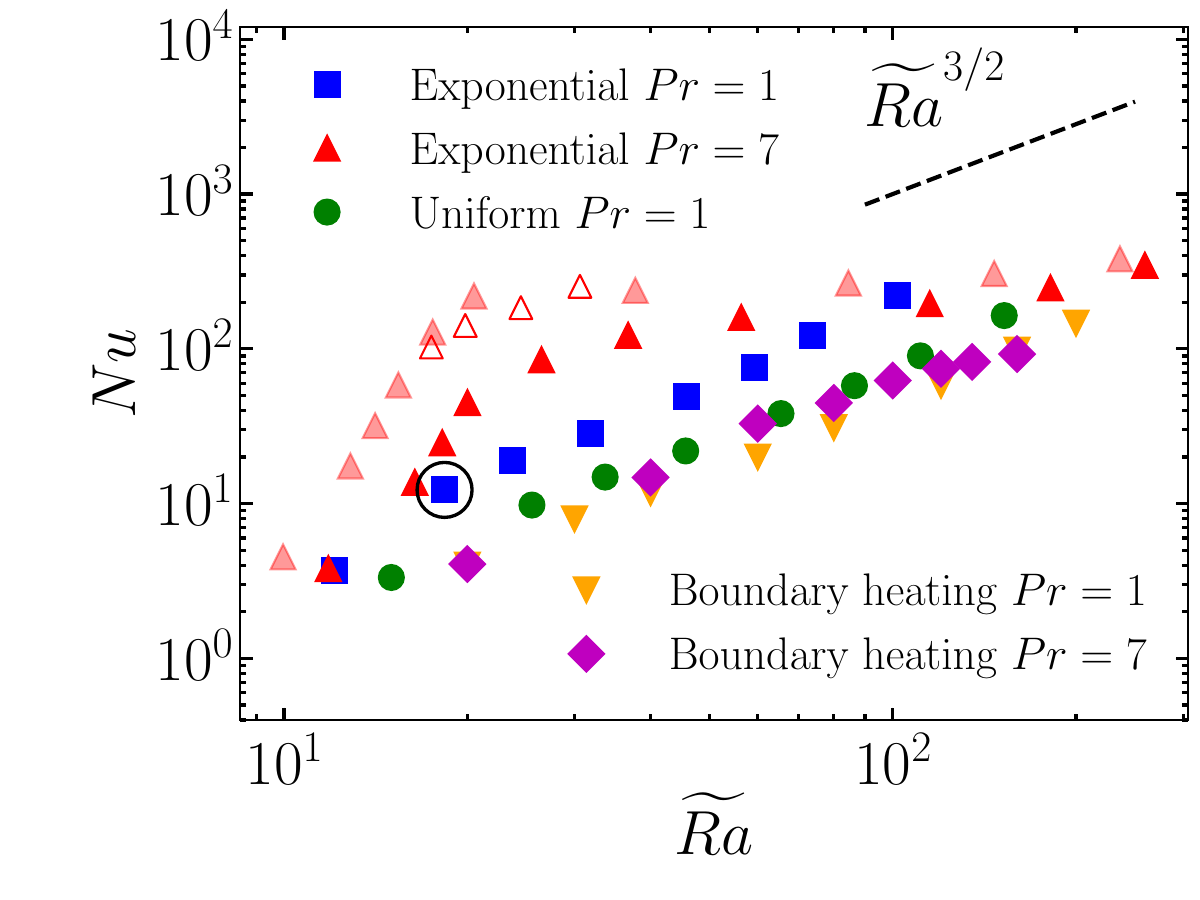}}
\subfloat[][]{\includegraphics[width=0.5\textwidth]{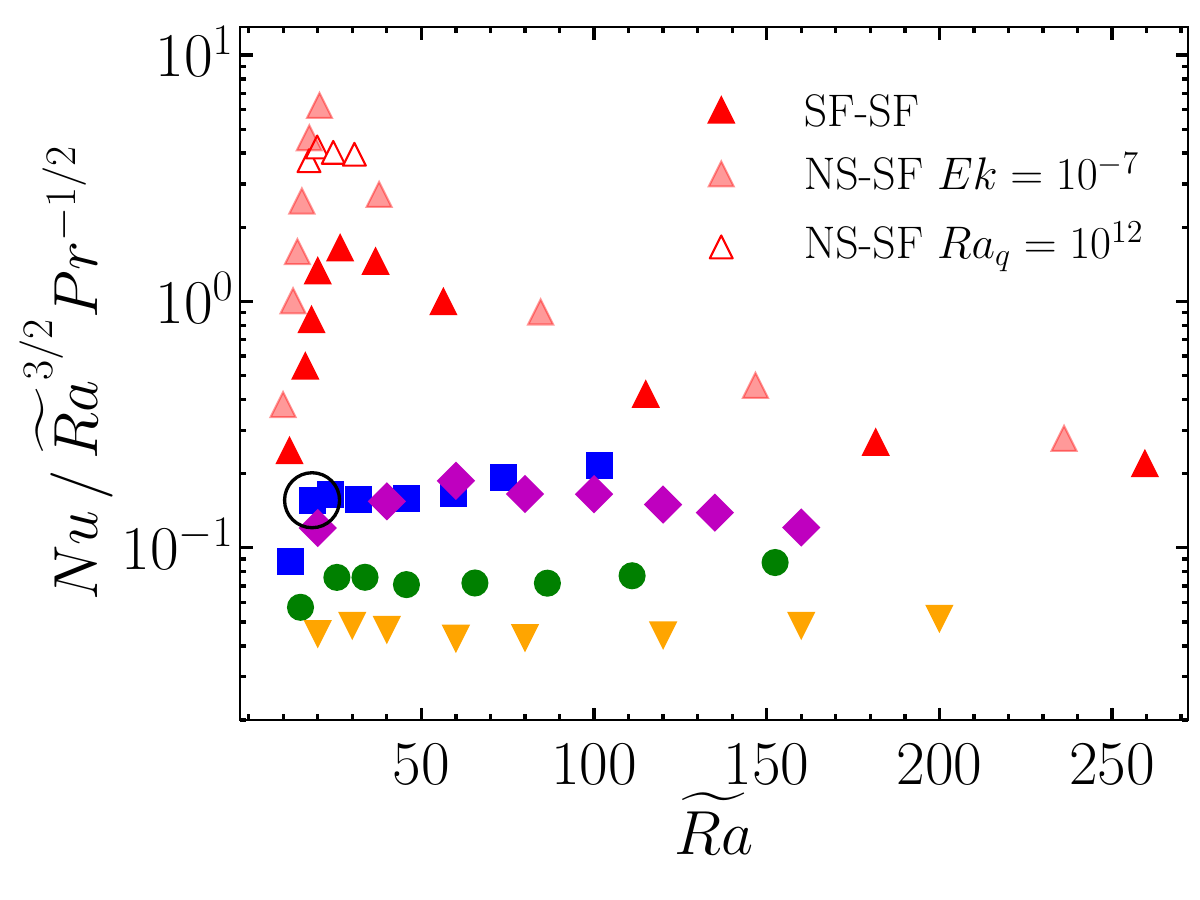}}\\
\subfloat[][]{\includegraphics[width=0.5\textwidth]{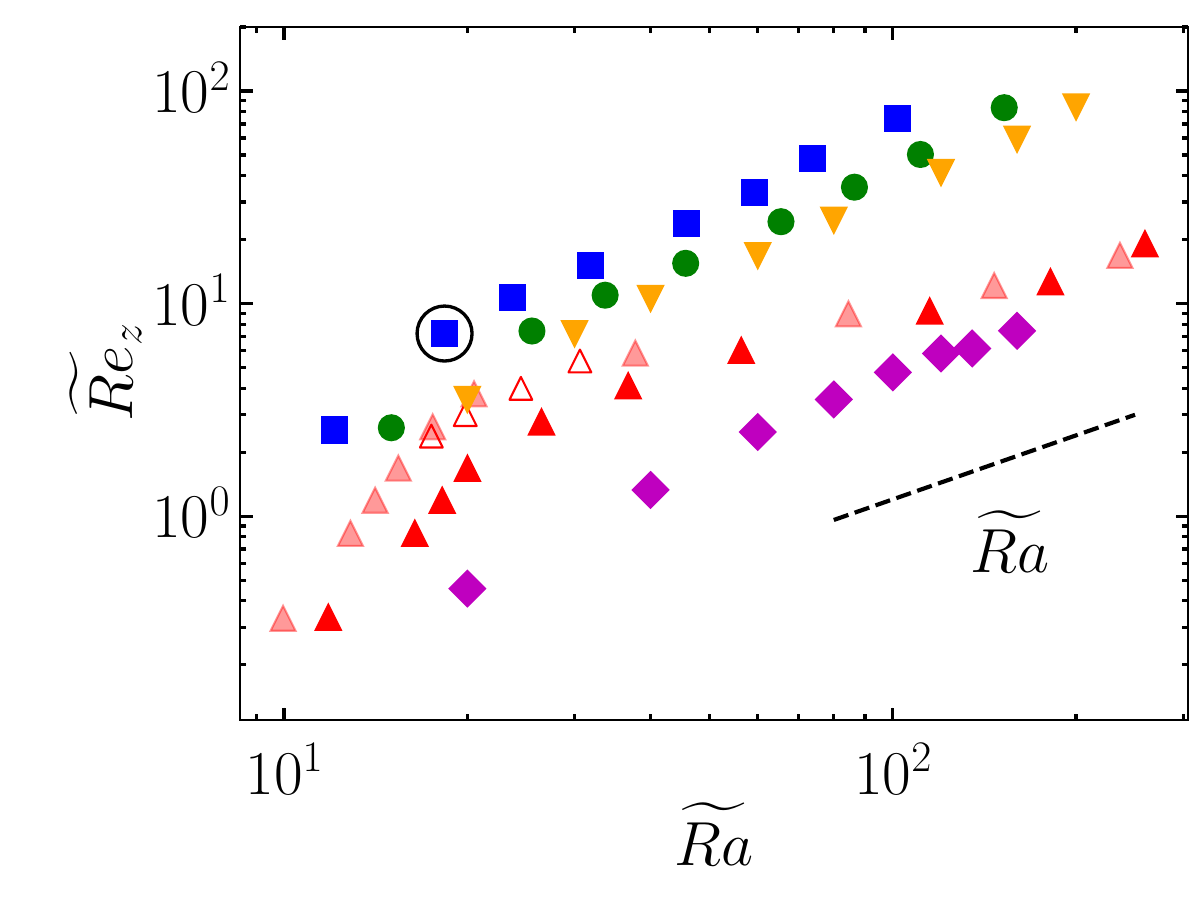}}
\subfloat[][]{\includegraphics[width=0.5\textwidth]{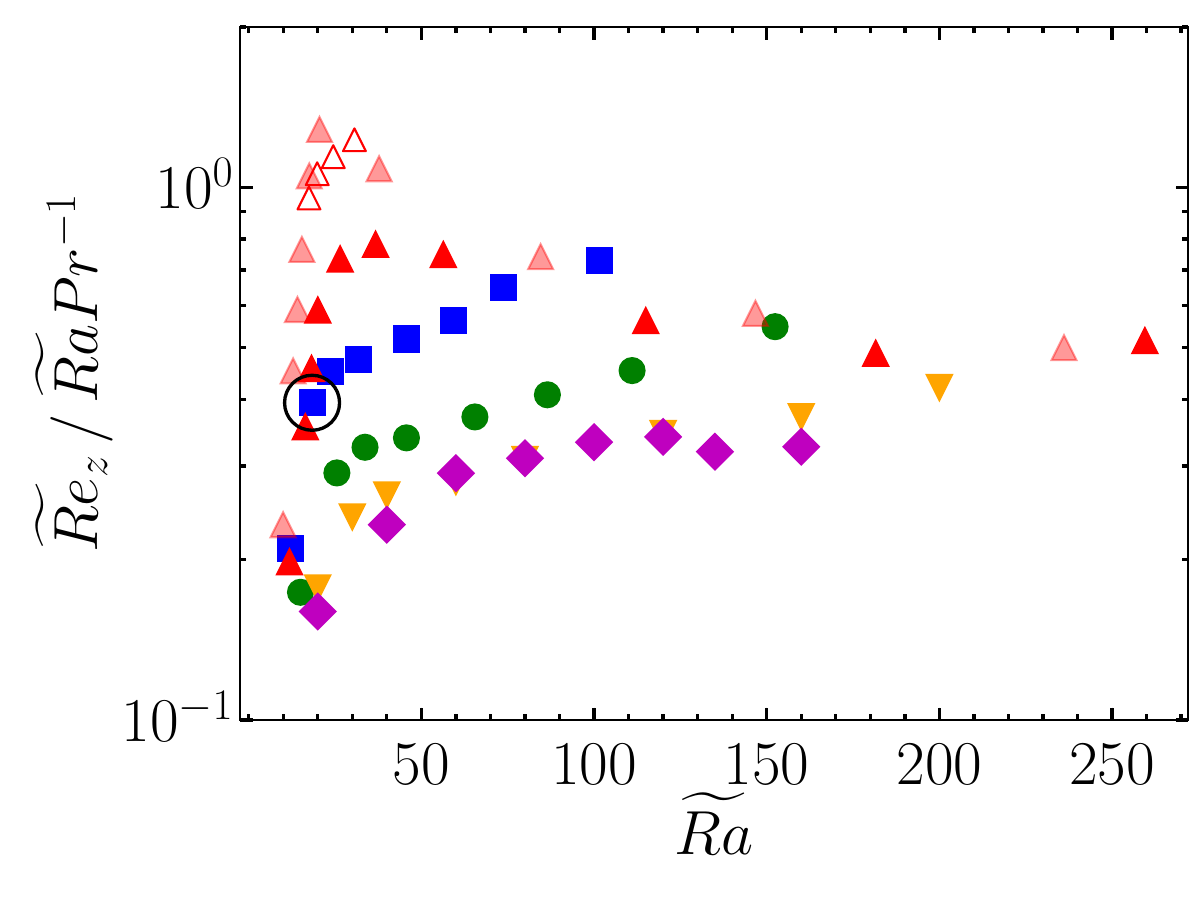}}\\
\caption{(a) Nusselt number, (b) compensated Nusselt number $Nu\,Pr^{1/2}/\Rat\vphantom{Ra}^{3/2}$, (c) vertical Reynolds number and (d) compensated vertical Reynolds number $\widetilde{Re}_z\,Pr/\Rat$. SF-SF cases are blue squares (Exponential, $Pr=1$), green circles (Uniform, $Pr=1$) and filled red triangles (Exponential, $Pr=7$); NS-SF cases are pale filled red triangles (fixed $Ek$) and open red triangles (matched to \cite{gH24}). Orange down-triangles and magenta diamonds are the boundary-heated data of \cite{sM21} at $Pr=1$ and $7$. The dashed lines in (a) and (c) show the diffusion-free slopes of \eqref{e:df}. The circled point is discussed in the text.}
\label{F:Nu}
\end{center}
\end{figure}

\begin{table}
\centering
\begin{tabular}{l c c c c c c}
\hline
\hline
Series & $Pr$ & $\Rat$ range & $\beta_{Nu}$ & $C_{Nu}$ & $\beta_{Re}$ & $C_{Re}$ \\
\hline
Exponential      & 1 & 46--102 & 1.90 & 0.183 & 1.44 & 0.62 \\
Uniform          & 1 & 46--152 & 1.66 & 0.076 & 1.39 & 0.42 \\
Boundary heating & 1 & 40--200 & 1.57 & 0.046 & 1.28 & 0.33 \\
Exponential      & 7 & 56--260 & 0.49 & 0.48  & 0.73 & 0.58 \\
Boundary heating & 7 & 40--160 & 1.31 & 0.155 & 1.23 & 0.31 \\
\hline
\hline
\end{tabular}
\caption{Least-squares exponents $\beta$ of $Nu \sim \Rat\vphantom{Ra}^{\beta_{Nu}}$ and $\widetilde{Re}_z \sim \Rat\vphantom{Ra}^{\beta_{Re}}$, and the prefactors $C_{Nu} = \langle Nu\,Pr^{1/2}/\Rat\vphantom{Ra}^{3/2}\rangle$ and $C_{Re} = \langle \widetilde{Re}_z\,Pr/\Rat\rangle$ averaged over the cases with $\Rat \ge 40$, for the SF-SF series of figure~\ref{F:Nu} and the boundary-heated data of \cite{sM21}. The diffusion-free prediction \eqref{e:df} is $\beta_{Nu} = 3/2$ and $\beta_{Re} = 1$ with $C_{Nu}$ and $C_{Re}$ independent of $\Rat$ and $Pr$.}
\label{T:prefactors}
\end{table}

Figure~\ref{F:Nu} shows the Nusselt and vertical Reynolds numbers, together with boundary-heated reference data \citep{sM21}; table~\ref{T:prefactors} lists the fitted exponents and prefactors beyond the near-onset transitions, $\Rat \ge 40$. At $Pr=1$ the Uniform and Exponential cases follow the expected $Nu \sim \Rat^{3/2}$ scaling over a wide range; the compensated Exponential data rise by $30\%$ between $\Rat = 59$ and $102$, whereas the compensated Uniform data vary by less than $10\%$ between $\Rat = 25$ and $111$ and rise by a further $13\%$ at $\Rat = 152$, the case with the shortest averaging interval. The two profiles do not, however, share a common prefactor: at essentially the same reduced Rayleigh number, $\Rat = 45.7$, the Exponential and Uniform profiles give $Nu = 49.2$ and $21.9$ respectively, a factor of $2.2$ at identical $\Rat$ and $Pr$, of which the differing normalisation constants account for only $10\%$. The Uniform data, meanwhile, track the boundary-heated $Pr=1$ data with a nearly constant offset, $Nu \approx 1.6\,Nu_{\mathrm{BH}}$ at fixed $\Rat$ across the full range of overlap, so the three forcings share an exponent while their prefactors span a factor of $4$ (table~\ref{T:prefactors}). For boundary heating at $Pr=1$ the prefactor, $C_{Nu} = 0.046$, is within $15\%$ of the value $1/25$ obtained by \cite{kJ12b} from the same reduced model, which confirms that the comparison is made on a common footing. The prefactor depends on the forcing and the geometry, in agreement with \cite{vB21}. The prefactor in $Nu = C\,\Rat^{3/2}Pr^{-1/2}$ therefore depends on the form of the heating even where the exponent is well satisfied, so the exponent alone does not fix the heat transport. 

We observe a significant difference in heat transport between the heating profiles: at $Pr=1$ the Uniform case gives markedly smaller Nusselt numbers than the Exponential case at comparable $\Rat$. The difference lies in $\Delta T$ rather than in the vigour of the flow. At the same $\Rat_q$ the two profiles give the same $\widetilde{Re}_z$ to within a few percent. This result might be expected since the depth-integrated buoyancy flux $\int_0^1 F\,dZ$ differs by only $10\%$. In contrast, $\Delta T$ is $40$--$50\%$ larger for the Uniform profile. With Uniform heating the flux $F(Z) = Z$ vanishes at the insulating lower boundary, so the mean temperature rises only slowly from its bottom value and the temperature drop is concentrated in the upper part of the layer, where the flux is largest and must be conducted through the top boundary. With Exponential heating the flux is deposited within $Z \lesssim 0.05$ of the base and both boundaries are insulating, so the drop is spread across the interior and no conductive layer forms (figure~\ref{F:mean_thermal_profiles}). The Exponential profile therefore reaches a given $\Delta T$, and hence a given $\Rat$, at a larger $\Rat_q$, $5\times10^{3}$ against $2\times10^{3}$ at $\Rat = 45.7$, with correspondingly larger $Nu$ and $\widetilde{Re}_z$.

At $Pr = 7$ the diffusion-free scaling of $Nu$ fails over most of the range surveyed. There is roughly a factor of ten separating the compensated Exponential data at $Pr=1$ and $Pr=7$ near $\Rat \sim 40$, so the $Pr^{-1/2}$ scaling does not collapse the data. Over $12 \le \Rat \le 260$ the compensated $Pr=7$ data rise by a factor of $\approx 7$ to a maximum near $\Rat \approx 26$ and then fall by a factor of $\approx 8$, and the boundary-heated $Pr=7$ data of \cite{sM21} behave in the same way over $20 \le \Rat \le 160$, so at this Prandtl number internal heating neither improves nor worsens the approach to the scaling. The separation is largest near the maximum of the $Pr=7$ compensated data and narrows as $\Rat$ increases, so whether the $Pr^{-1/2}$ collapse is recovered asymptotically cannot be settled with the present range. A comparably strong $Pr$ dependence was found by \cite{bM20} in non-rotating convection driven by heat sources and sinks.

The vertical Reynolds number is further from the diffusion-free prediction. At $Pr=1$ all internally heated cases scale more steeply than $\Rat^{1}$, as do the boundary-driven cases. At $Pr=7$ the compensated data rise to a maximum near $\Rat \approx 40$ and then decrease; only the two largest $\Rat$ cases show any flattening, and these are precisely the cases whose $Nu$ scales more weakly than $\Rat^{3/2}$. The $Pr=1$ data thus show the behaviour reported by \cite{vK25b} for boundary-heated stress-free convection, in which the heat transport reaches its diffusion-free scaling well before the momentum transport does, and by the criterion of \cite{jS24}, which requires both, none of our cases is in the GT regime. If a diffusion-free regime exists, larger $\Rat$ than tested here is needed to reach it.

\begin{figure}
\begin{center}
\subfloat[][]{\includegraphics[width=0.3\textwidth]{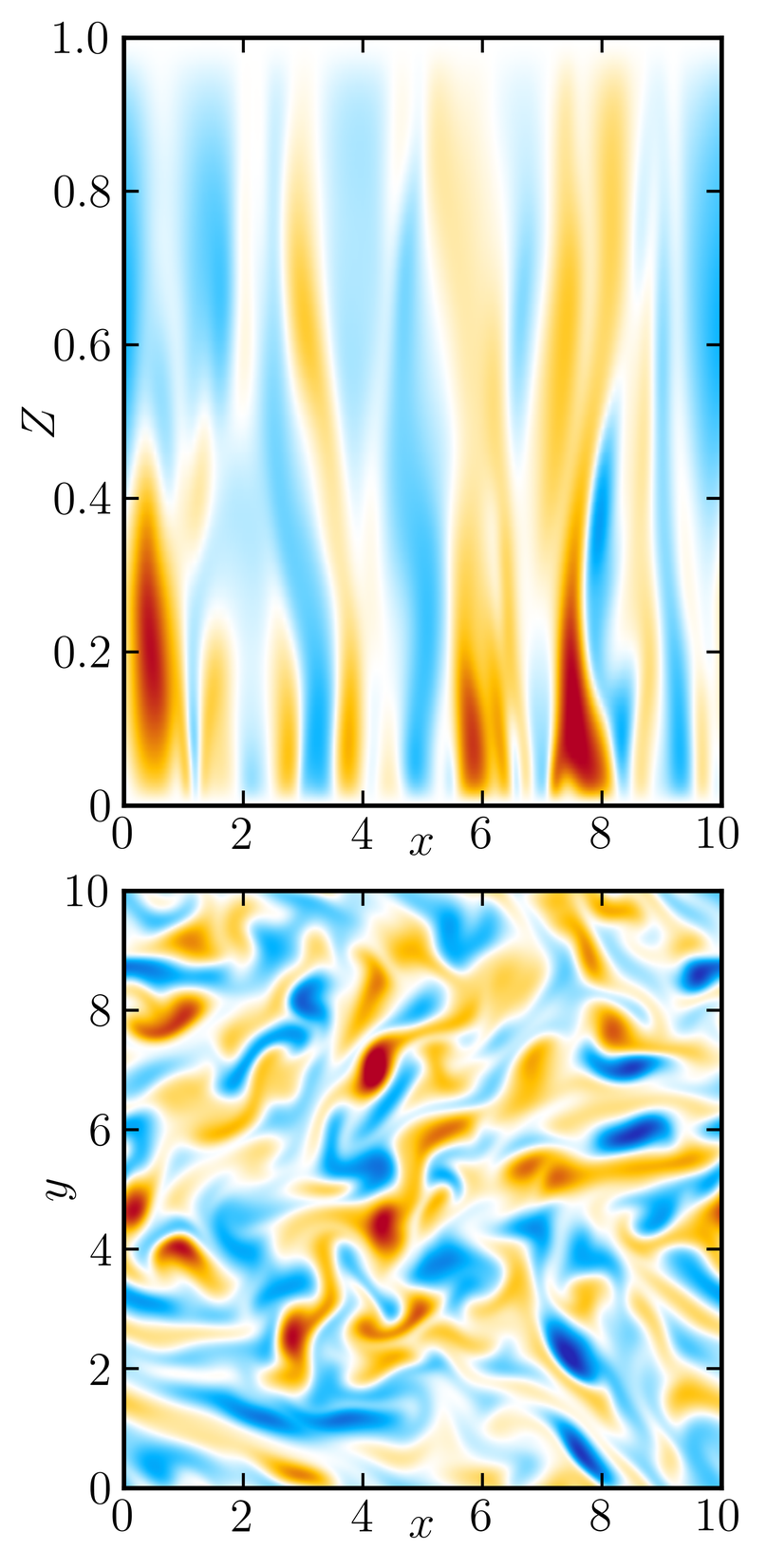}}%
\hspace{0.06\textwidth}%
\subfloat[][]{\includegraphics[width=0.3\textwidth]{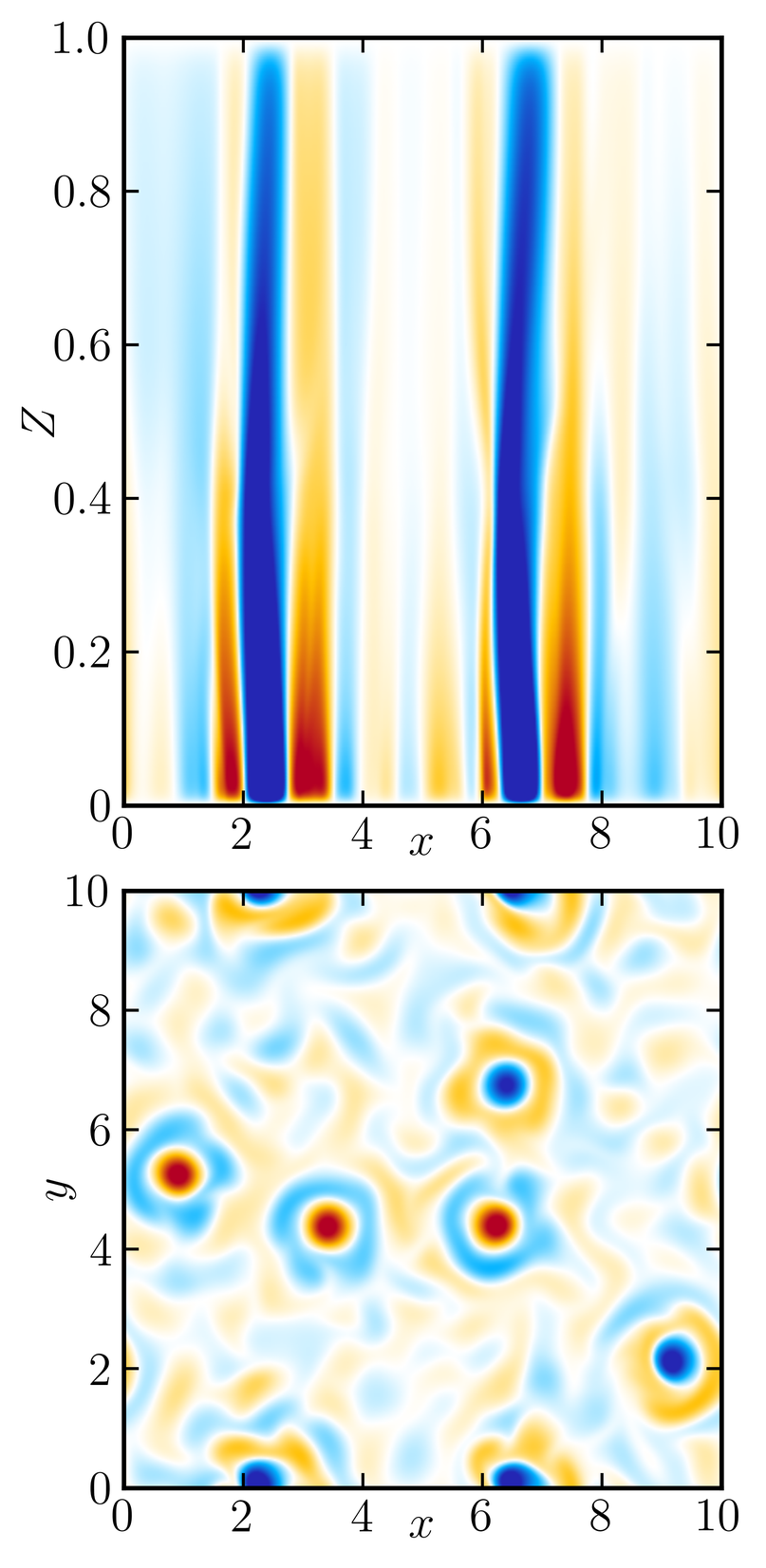}}\\
\caption{Fluctuating temperature for $Q(Z) = \mathcal{E}_c(0.05,Z)$ at nearly equal reduced Rayleigh number: (a) $Pr=1$, $\Rat = 18.4$ (the circled point in figure~\ref{F:Nu}) and (b) $Pr=7$, $\Rat = 20.0$. Within each panel the upper plot is an $xz$ slice and the lower an $xy$ slice taken at the mid-plane, $Z=0.5$, with $x$ and $y$ in units of $\lambda_c$, so one unit of $x$ or $y$ is one critical wavelength and the domain is ten wavelengths wide. Red (blue) denotes positive (negative) fluctuating temperature.}
\label{F:vis}
\end{center}
\end{figure}

The apparent collapse of the $Pr=1$ data is already present at modest $\Rat$, and this is informative. Consider the circled point in figure~\ref{F:Nu}, at $\Rat = 18.4$. A scaling analysis would place this case in the diffusion-free regime, yet the visualisation in figure~\ref{F:vis}(a) shows a relatively laminar flow whose horizontal scale is set by the viscously controlled critical wavelength, as is to be expected this close to onset, $\Rat_c \approx 16.6$ for this profile. Scaling collapse alone is therefore not sufficient evidence of diffusion-free dynamics: laminar and weakly turbulent flows can sit on the same scaling line. Figure~\ref{F:vis} also shows that at nearly the same $\Rat$ the $Pr=7$ flow retains coherent columns spanning the full depth and a horizontal cross-section of isolated convective Taylor columns.

\subsection{Comparison with radiatively driven DNS}
\label{S:comparison}

\begin{figure}[t]
\begin{center}
\subfloat[][]{\includegraphics[width=0.5\textwidth]{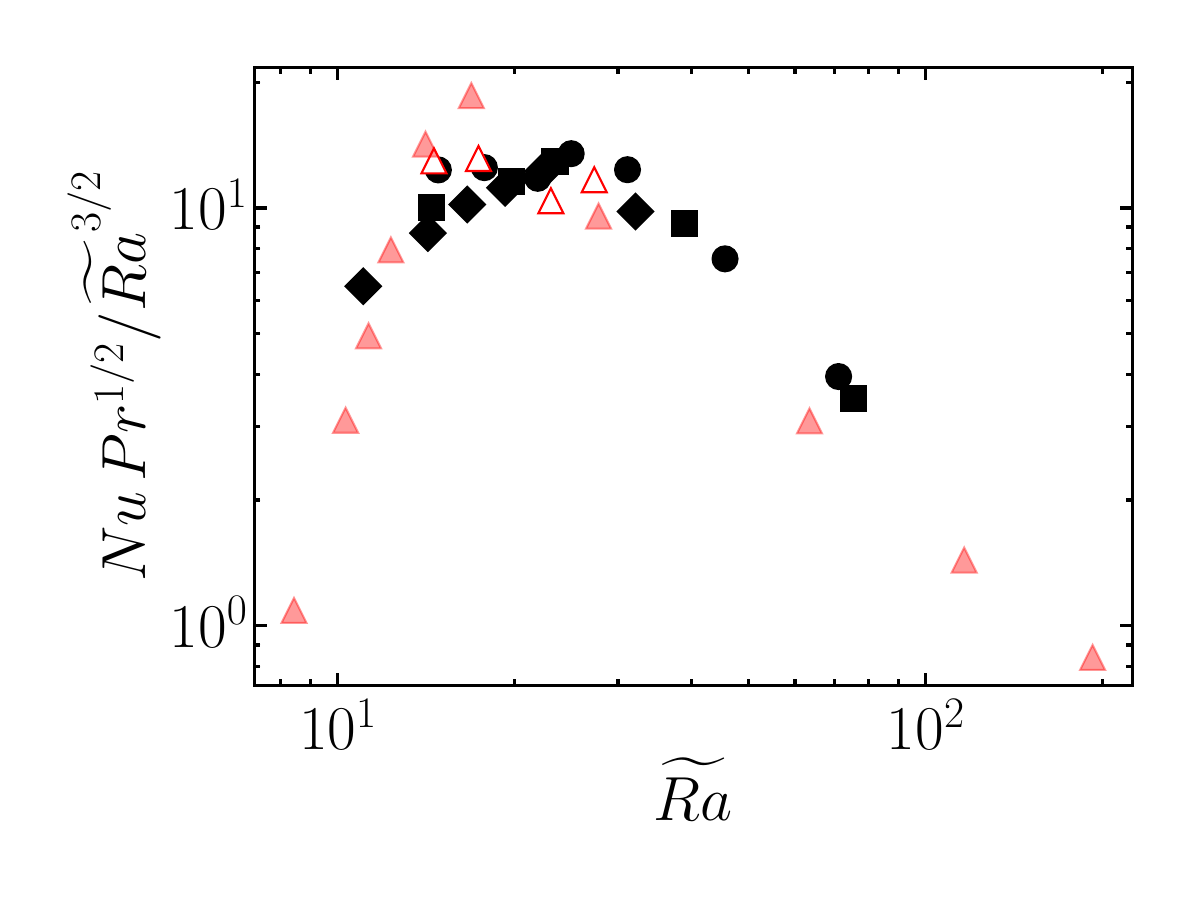}}
\subfloat[][]{\includegraphics[width=0.5\textwidth]{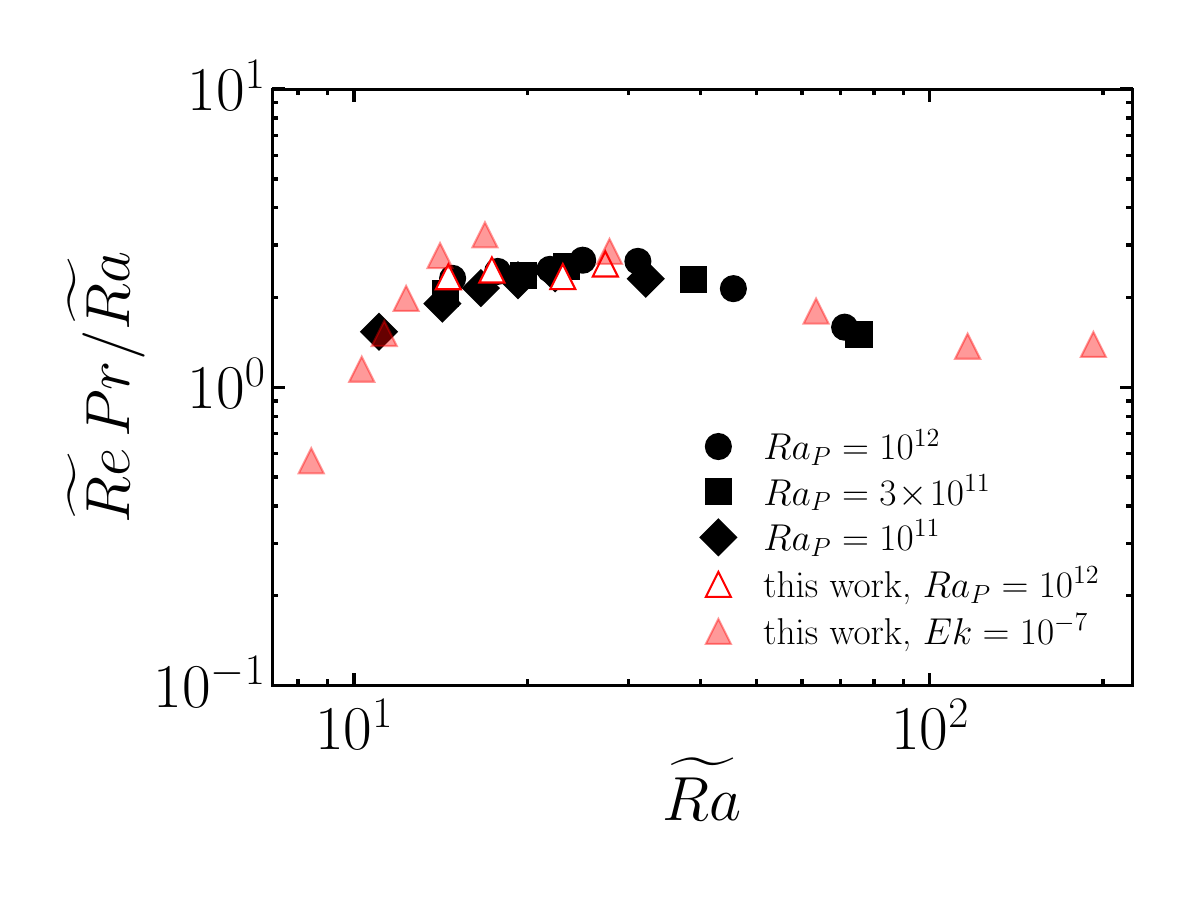}}\\
\caption{Compensated (a) $Nu$ and (b) $\widetilde{Re}$ for our NS-SF $Pr=7$ cases (red) and the DNS of \cite{gH24} (black). Open red triangles are matched to their $Ra_q = 10^{12}$ series and use the same Ekman numbers; pale filled red triangles are the broader scan at fixed $Ek = 10^{-7}$. Black symbols encode the DNS series at $Ra_q = 10^{12}$ (circles), $3\times10^{11}$ (squares) and $10^{11}$ (diamonds); the Ekman number varies along each. Here $Nu$ and $\Rat$ use their normalisation, with $\Delta T$ taken between $Z=0$ and $Z=0.75$, and so differ from the values tabulated in table~\ref{T:data}; $\widetilde{Re}$ is formed from all three velocity components.}
\label{F:dns}
\end{center}
\end{figure}

Figure~\ref{F:dns} compares our NS-SF $Pr=7$ cases with the DNS of \cite{gH24}. The matched cases are computed at the same flux-based Rayleigh and Ekman numbers as their $Ra_q = 10^{12}$ series and use the same definition of $\Delta T$. Because the heat flux is imposed, $\Rat = \Delta T\,\Rat_q$ is an output and differs slightly between the two models even at identical control parameters, which is why the open triangles do not sit at exactly the abscissae of the circles. With these conventions they track the DNS across the range sampled, validating the reduced Ekman-pumping model against the radiatively driven simulations; the agreement is closest at the smallest $\Rat$, with our values falling modestly below the DNS as $\Rat$ increases.

Our scan at fixed $Ek = 10^{-7}$ overlaps the same range of $\Rat$; compared at the same flux Rayleigh number it agrees with the matched cases to within the $10$--$20\%$ time-averaging uncertainty in $Nu$ and to within $4\%$ in $\widetilde{Re}$, for $Ek$ between $10^{-7}$ and $1.5\times10^{-6}$, so the reduced model reproduces the $Ek$-independence found by \cite{gH24} over the range they cover.

The compensated Nusselt number does not saturate. In the DNS it reaches a maximum near $\Rat \approx 25$ and has fallen by a factor of about $3.5$ by $\Rat \approx 75$; along our fixed-$Ek$ scan it peaks near $\Rat \approx 17$ and falls by a factor of $22$ by $\Rat = 192$. Beyond $\Rat \approx 30$ the DNS and the reduced model fall at the same rate, the DNS lying a factor of about $1.5$ higher, consistent with the stronger Ekman pumping at their larger $Ek$. \cite{gH24} place the DNS points beyond $\mathcal{R} = Ra_P Ek^{3}/Pr^{2} = 3\times10^{-7}$ outside their GT regime, but the reduced model, which enforces geostrophy at leading order, shows the same downturn, as do the SF-SF data on the $Ek = 0$ manifold (\S\ref{S:scalings}). The downturn is therefore a property of the geostrophic dynamics, and the approximately constant compensated $Nu$ over the range sampled by the DNS is the neighbourhood of a maximum rather than a diffusion-free regime. The compensated $\widetilde{Re}$ (figure~\ref{F:dns}b) behaves in the same way with a shallower maximum.

The experiments of \cite{vB21} vary both the rotation rate and the layer depth, and together with the DNS of \cite{gH24} they cover a decade in each of $Ek$ and $Ra_P$; the collapse of these data in diffusion-free variables is a genuine test of $Ek$-independence over that range, and, as shown above, the reduced model reproduces it. What the data do not test is the $Pr$ dependence, which was fixed at that of the working fluid in both the experiments and the DNS; and the DNS points that do extend beyond $\Rat \approx 30$ show the start of the same downturn as the reduced model. Establishing the asymptotic regime requires both: the predicted $Pr^{-1/2}$ collapse fails in the reduced model (\S\ref{S:scalings}), and the compensated transport falls by more than an order of magnitude beyond the sampled range. We therefore do not dispute the data or their collapse over the range sampled; we dispute the interpretation of that collapse as diffusion-free behaviour, since the range sampled is not in the GT regime.

\subsection{Signatures of geostrophic turbulence}
\label{S:gt}

\begin{figure}
\begin{center}
\subfloat[][]{\includegraphics[width=0.45\textwidth]{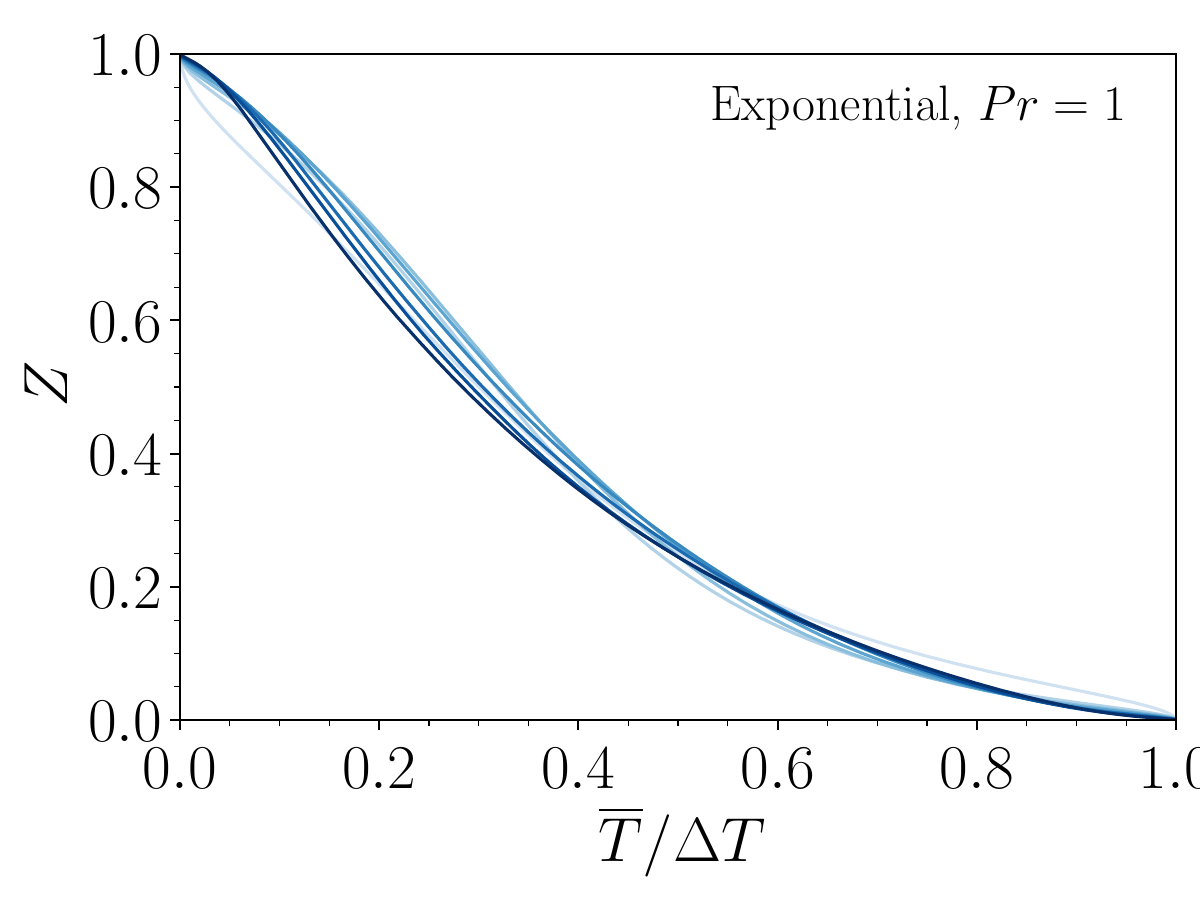}}
\subfloat[][]{\includegraphics[width=0.45\textwidth]{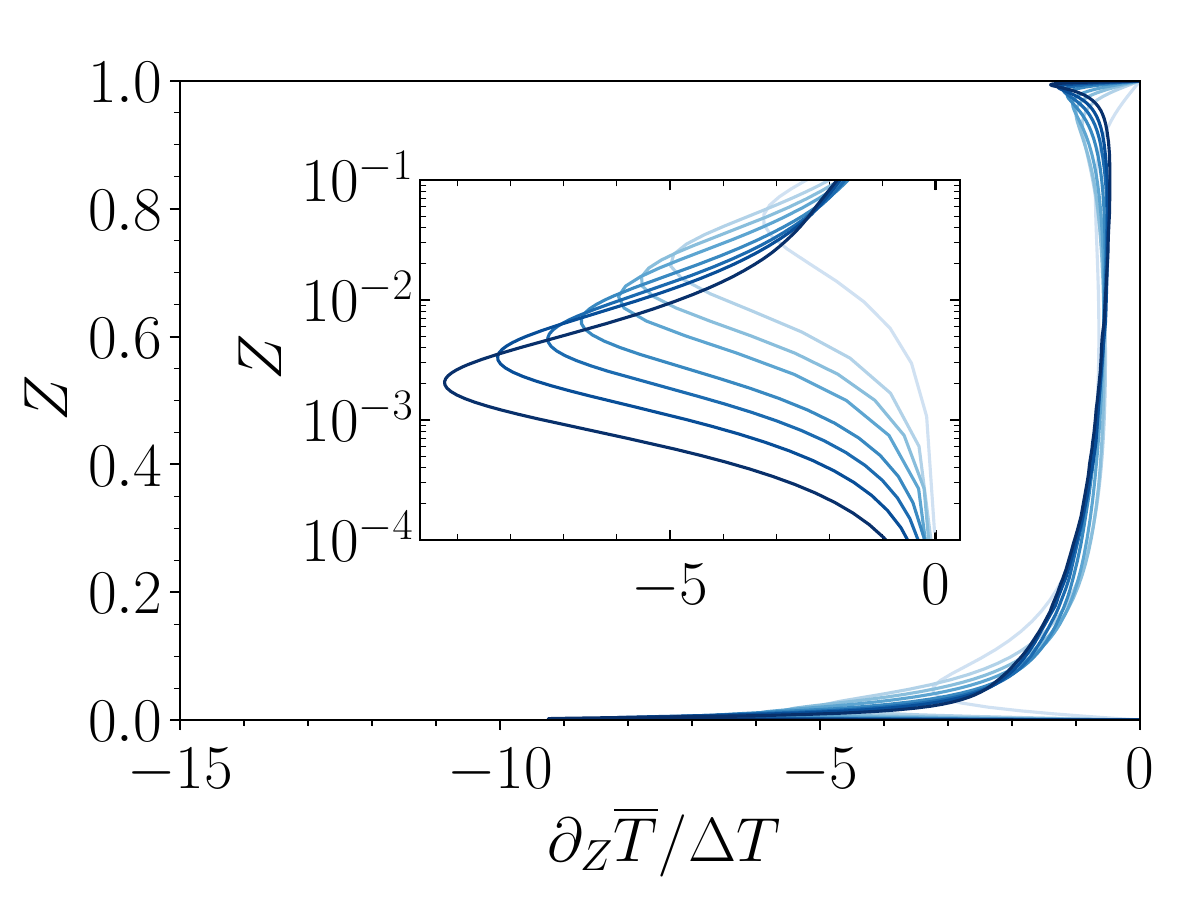}}\\
\subfloat[][]{\includegraphics[width=0.45\textwidth]{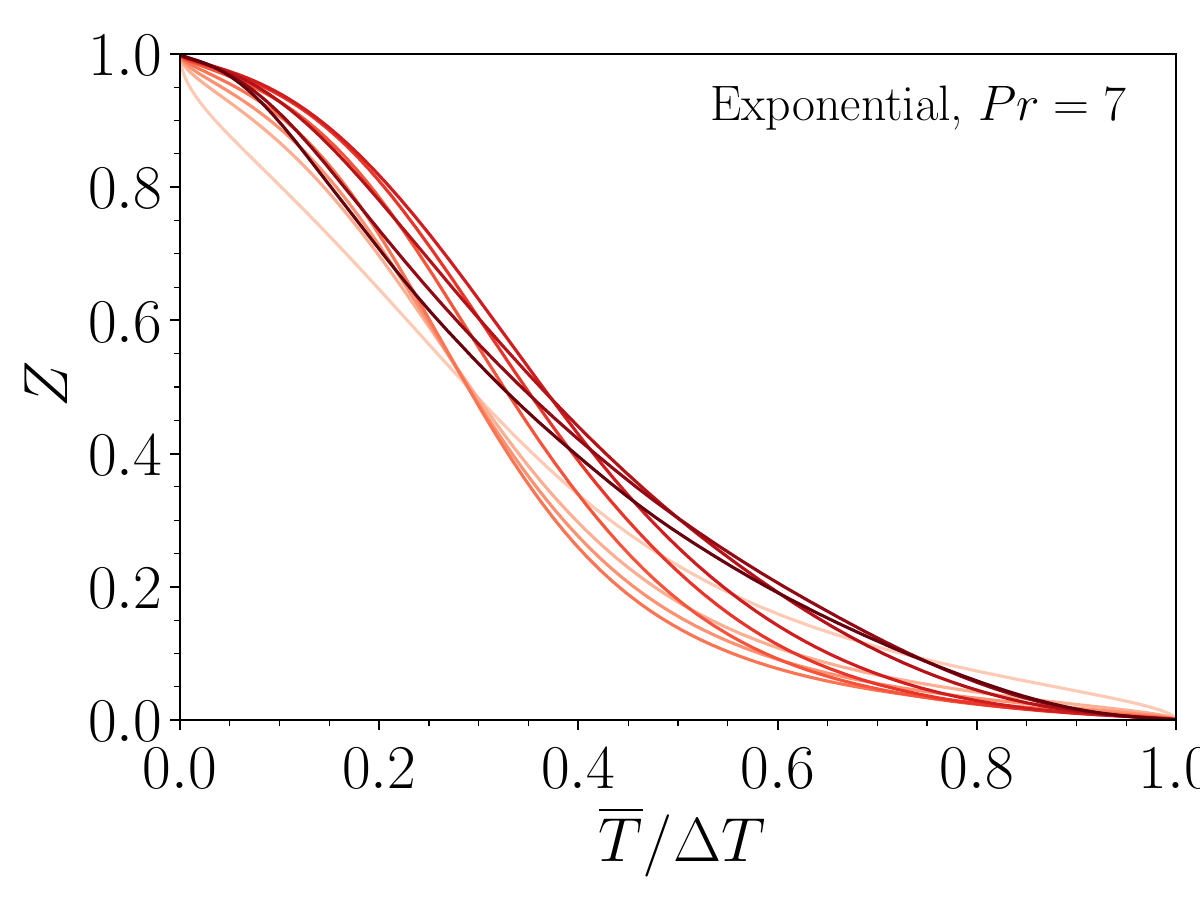}}
\subfloat[][]{\includegraphics[width=0.45\textwidth]{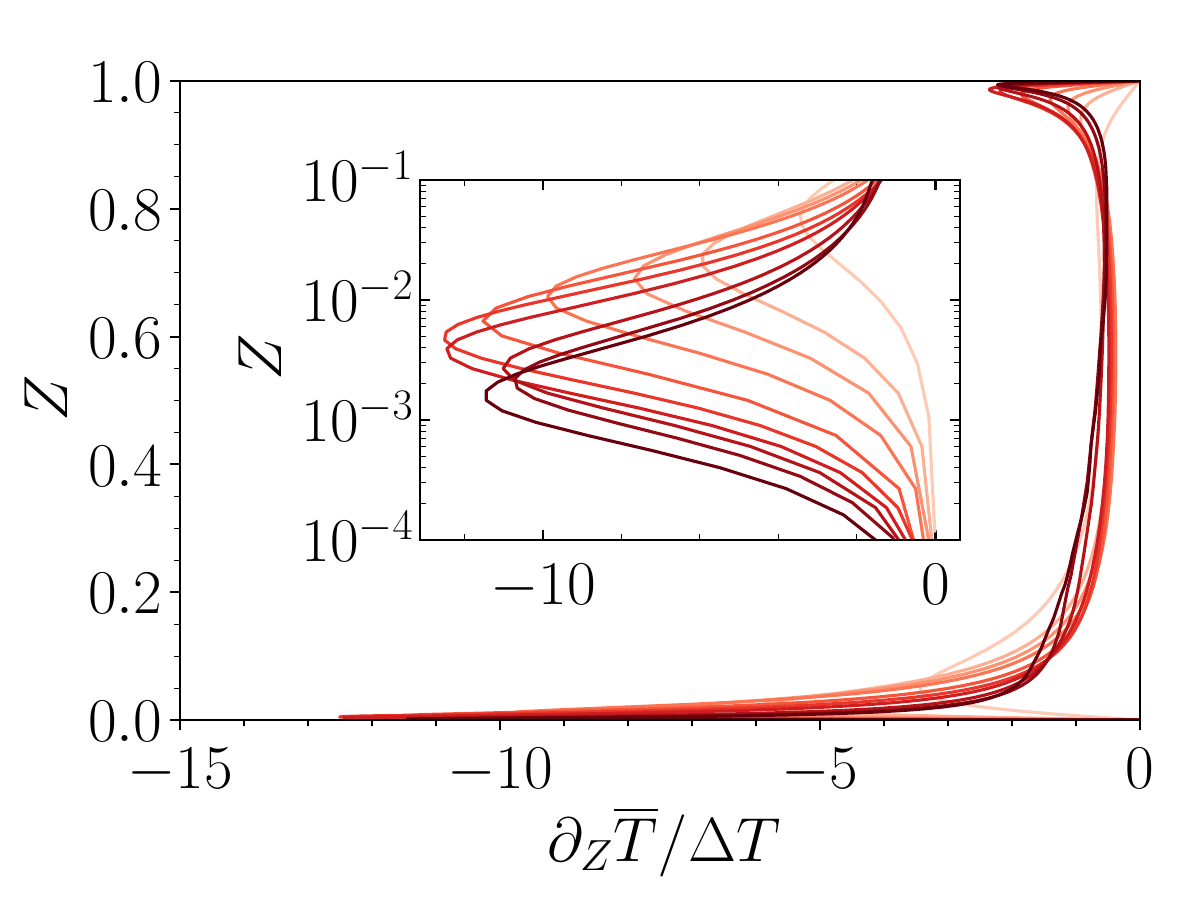}}\\
\subfloat[][]{\includegraphics[width=0.45\textwidth]{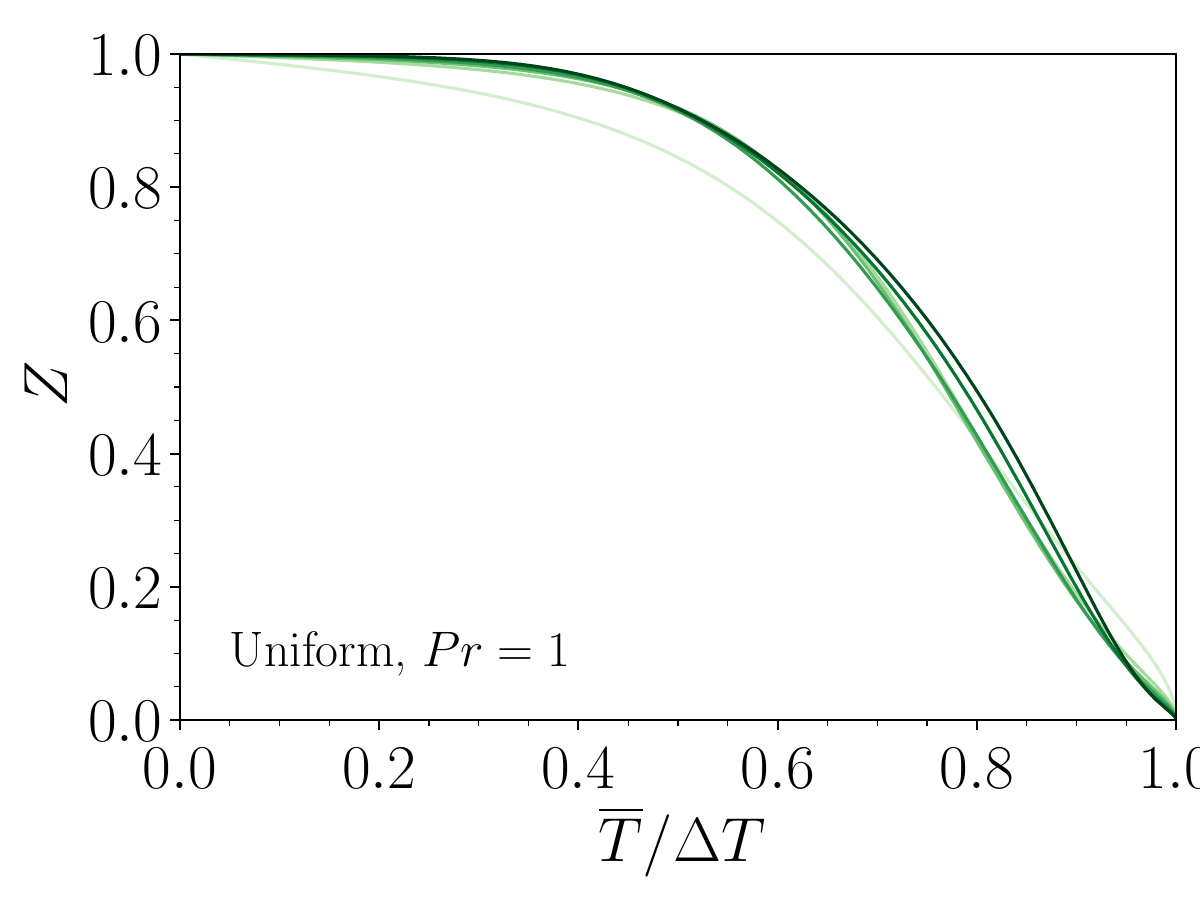}}
\subfloat[][]{\includegraphics[width=0.45\textwidth]{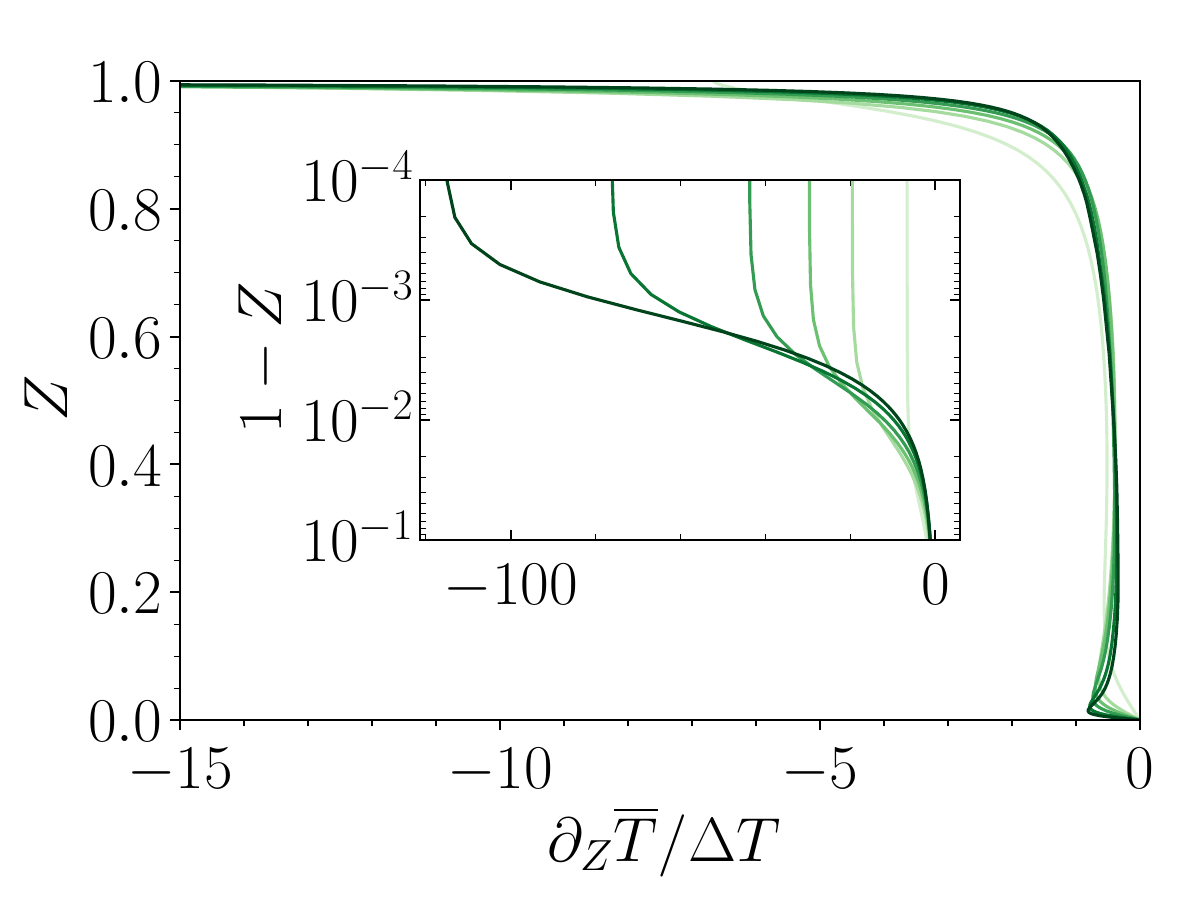}}\\
\caption{Profiles of the time-averaged mean temperature (left column) and its vertical gradient (right column), both normalised by $\Delta T$, for the SF-SF cases: Exponential profile at $Pr=1$ (a,b; blue), Exponential at $Pr=7$ (c,d; red) and Uniform at $Pr=1$ (e,f; green). Darker colours correspond to larger $\Rat$. The insets show the gradient near the heated lower boundary (b,d) and near the cooled upper boundary (f) on a logarithmic scale.}
\label{F:mean_thermal_profiles}
\end{center}
\end{figure}

\begin{figure}
\begin{center}
\subfloat[][]{\includegraphics[width=0.5\textwidth]{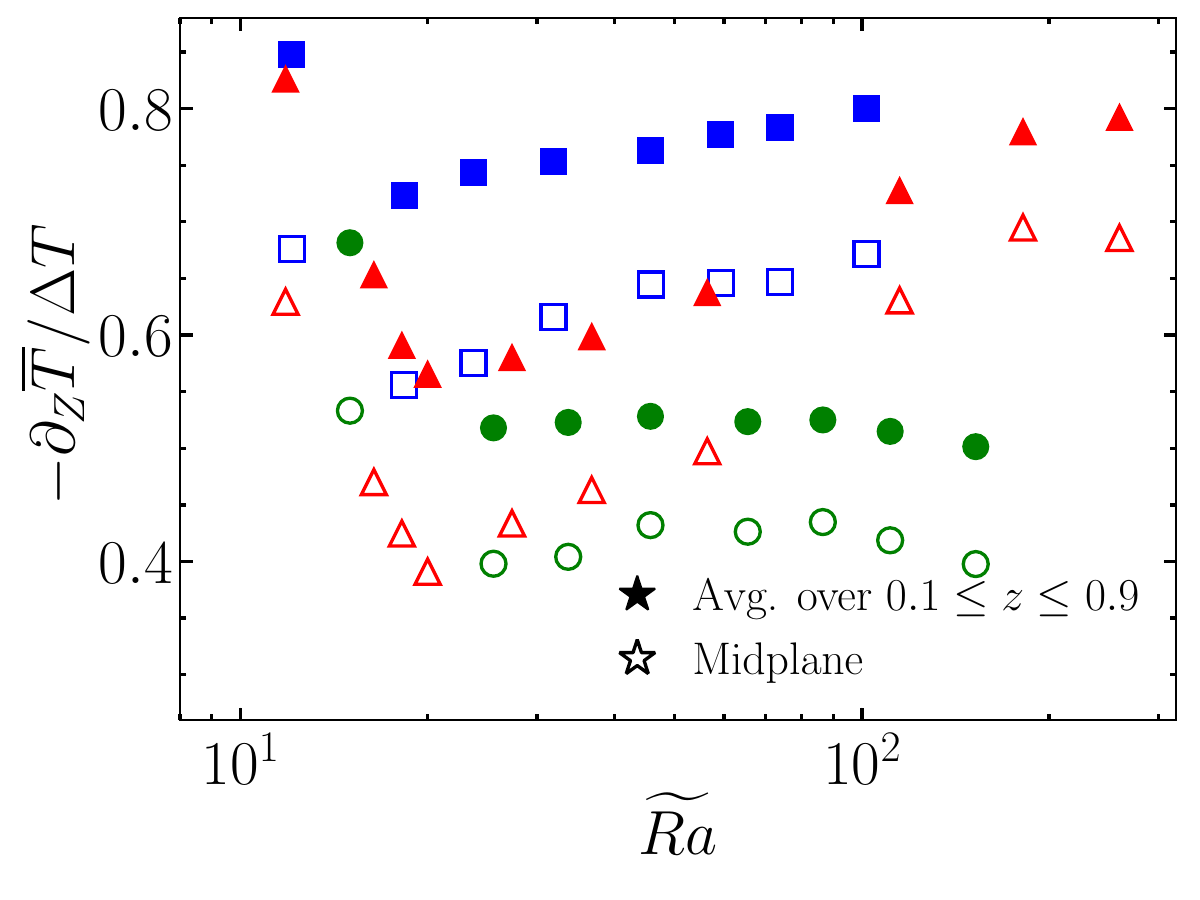}}
\subfloat[][]{\includegraphics[width=0.5\textwidth]{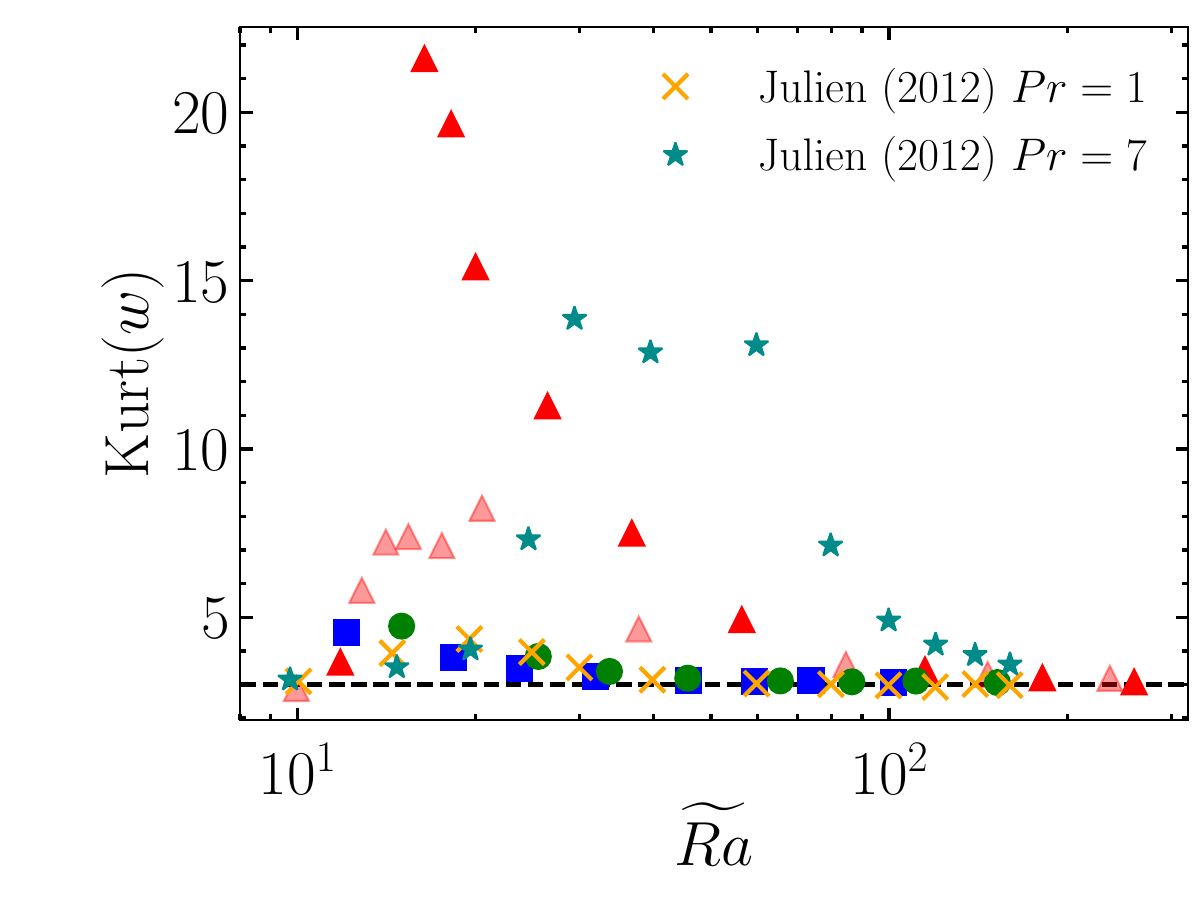}}\\
\caption{(a) Vertical average over $0.1 \le Z \le 0.9$ (filled symbols) and mid-plane value (open symbols) of the mean temperature gradient, normalised by $\Delta T$; only SF-SF cases appear in this panel. (b) Kurtosis of the vertical velocity at the mid-plane; the dashed line marks the Gaussian value of $3$. Colours and shapes follow figure~\ref{F:Nu}: blue squares Exponential $Pr=1$, green circles Uniform $Pr=1$, red triangles Exponential $Pr=7$, and pale filled triangles the NS-SF cases. Stars and crosses are the boundary-heated data of \cite{kJ12} at $Pr=7$ and $Pr=1$ respectively.}
\label{F:gt}
\end{center}
\end{figure}

We now turn to the two criteria of \cite{kJ12}, which characterise the interior dynamics directly rather than through a bulk transport law. The first is the horizontally averaged temperature gradient in the interior, $\dZ\overline{T}/\Delta T$. As the flow becomes turbulent this gradient saturates at an $O(1)$ value, so that a fixed fraction of the temperature drop is maintained across the interior, and this saturation marks the point beyond which the interior transport mechanism no longer changes with $\Rat$. The second criterion is the kurtosis of the vertical velocity, $\mathrm{Kurt}(w) = \langle w^4\rangle/\langle w^2\rangle^2$. A flow of sparse, intense plumes has a strongly non-Gaussian $w$ with $\mathrm{Kurt}(w) \gg 3$; as the plumes break down into turbulence the distribution becomes Gaussian and the kurtosis decays to $3$. Neither quantity involves an undetermined prefactor, and GT is identified \textit{empirically} where both have saturated with increasing $\Rat$.

Figure~\ref{F:mean_thermal_profiles} shows the time-averaged mean temperature and its vertical gradient, normalised by $\Delta T$, for the three SF-SF series. For the Exponential profile the temperature drop is distributed across the depth and the gradient is $O(1)$ throughout the interior; the only sharp feature is at $Z = 0$, where the heat is deposited, and the inset shows that this layer thins as $\Rat$ increases, non-monotonically at $Pr = 7$. For the Uniform profile the mean temperature varies more weakly over the lower half of the layer, where the flux $F(Z) = Z$ is small, and the drop is concentrated in the upper part. The conductive layer at $Z = 1$, through which the entire flux leaves, thins and steepens with increasing $\Rat$ (inset), in the manner of the thermal boundary layer of boundary-heated convection \citep{sM21}. This difference in the distribution of $\Delta T$ between the two profiles is the one invoked in \S\ref{S:scalings} to explain the smaller Nusselt number of the Uniform profile. In all three series the normalised profiles approach a fixed shape as $\Rat$ increases, with an interior gradient that is non-zero and $O(1)$; this is the saturation that the first criterion describes, and figure~\ref{F:gt}(a) quantifies it.

In contrast to the transport scalings, the interior mean temperature gradient of figure~\ref{F:gt}(a) is well behaved. As $\Rat$ increases, $\dZ\overline{T}/\Delta T$ settles onto an $O\lb 1\rb $ value that depends on the heating profile. The Uniform cases are flat to within a few percent for $\Rat \gtrsim 25$; the Exponential cases pass through a shallow minimum and then increase slowly, and are still rising weakly at the largest $\Rat$, most noticeably at $Pr = 7$, so we do not claim a strict plateau. The relevant comparison is with the compensated transport over the same range: for the Exponential $Pr=7$ cases the compensated Nusselt number falls by a factor of $7.5$ from its maximum near $\Rat \approx 26$ to its value at $\Rat = 260$, whereas the vertically averaged $\dZ\overline{T}/\Delta T$ varies by less than a factor of $1.5$ across the whole range and remains $O\lb 1\rb $ throughout. Similar behaviour is observed in other studies of rotating convection with fixed-temperature boundary heating \citep{mS06,tO23}. The interior gradient is therefore a far more stable diagnostic than the compensated $Nu$ or $\widetilde{Re}_z$.

The second criterion of \cite{kJ12}, the kurtosis of the vertical velocity, behaves in a similar way (figure~\ref{F:gt}b). At $Pr=1$ it falls towards the Gaussian value of $3$. At $Pr=7$ it first rises, from $3.7$ at $\Rat = 11.8$ to $21.6$ at $\Rat = 16.4$, and only then decays to $3.1$ at $\Rat = 260$, so saturation demands larger $\Rat$ at higher $Pr$ -- consistent with the slower approach of $\dZ\overline{T}/\Delta T$ in figure~\ref{F:gt}(a). The near-critical value is a caution in itself: the flow there is cellular, yet its kurtosis already lies near $3$, a flow of few smooth modes being unremarkably Gaussian. As with the scaling collapse of figure~\ref{F:Nu}, the criterion is diagnostic only well above onset. The Uniform cases satisfy both criteria most convincingly, but they do so no earlier than boundary-heated convection at the same $Pr$: over the common range $25 \lesssim \Rat \lesssim 120$ the compensated Nusselt number of figure~\ref{F:Nu}(b) varies by $10\%$ for the Uniform profile and by $13\%$ for the boundary-heated data. What distinguishes them is the prefactor, not the onset.

In the reduced model, GT is uniquely associated with high $\Rat$, and we conclude that both criteria of \cite{kJ12} remain reliable indicators of the regime when the convection is internally heated, over a range of forcing in which the transport scalings do not.

\section{Conclusion}
\label{S:conc}

We set out to test whether the diffusion-free regime of rapidly rotating convection is reached more easily when the system is heated internally rather than at the boundaries, and find that it is not. In both the doubly stress-free and the no-slip/stress-free configurations the scaling behaviour is not substantially different from the boundary-heated case. While the Nusselt number at $Pr=1$ appears to follow the diffusion-free $\Rat$ scaling, the vertical Reynolds number continues to increase more rapidly with $\Rat$ than \eqref{e:df} predicts, indicating that diffusion continues to play a role; similar behaviour was reported by \cite{vK25} and \cite{aK25}. The $Pr^{-1/2}$ scaling does not collapse the data at the two Prandtl numbers tested, and the prefactor depends on the heating profile.

Our results also bear on how such tests should be conducted. The presence of diffusion-free scalings in global transport quantities is insufficient to establish either a diffusion-free flow or geostrophic turbulence: the $Pr=1$ collapse persists into the visibly laminar regime (figure~\ref{F:vis}a), and the $Pr=7$ data pass through a local maximum in compensated $Nu$ over which any narrow sweep would appear diffusion-free. The reduced model with Ekman pumping reproduces the radiatively driven DNS of \cite{gH24} case by case, and shows that the range they sample sits at this maximum. Sampling a wide range of $\Rat$, and varying $Ek$ and $Pr$ independently, are both necessary. The asymptotic framework used here does this by construction, and finds clear residual $Pr$-dependence in $Nu$ and $\widetilde{Re}_z$, in both exponent and prefactor.

We do, however, confirm that the diagnostics of \cite{kJ12} continue to identify the transition to geostrophic turbulence in the internally heated system. The two criteria are complementary: the kurtosis of the vertical velocity supplies a universal threshold, its Gaussian value of $3$, independent of the heating profile and $Pr$, while the interior mean temperature gradient is a low-order statistic that converges rapidly under time averaging. We suggest that the saturation of these quantities, rather than the exponent of a bulk transport law, is the more accurate signature of the GT regime. Large-scale vortices may delay the approach of $\widetilde{Re}_z$ to its diffusion-free scaling, though previous work shows that even with the vortices explicitly removed the vertical Reynolds number does not reach it until $\Rat \gtrsim 200$ \citep{tO23}. 

\vspace{6pt}
\noindent\textbf{Supplementary data.} The simulation table of Appendix~\ref{A:table} is provided as a machine-readable CSV file.

\vspace{6pt}
\noindent\textbf{Acknowledgements.} Computations were performed on the Alpine supercomputer at the University of Colorado Boulder, and the Anvil supercomputer at Purdue University. Alpine is jointly funded by the University of Colorado Boulder, the University of Colorado Anschutz, and Colorado State University and with support from NSF grants OAC-2201538 and OAC-2322260. Anvil was made available through allocation PHY180013 from the Advanced Cyberinfrastructure Coordination Ecosystem: Services \& Support (ACCESS) program, which is supported by NSF grants 2138259, 2138286, 2138307, 2137603 and 2138296. The authors used Claude (Anthropic) during preparation of this manuscript for language editing and to verify reported quantities against the simulation output.

\vspace{6pt}
\noindent\textbf{Funding.} This work was supported by the National Science Foundation (NSF) through grant number EAR-1945270. S.M. acknowledges support from the European Research Council (agreement 833848-UEMHP, under the Horizon 2020 program).

\vspace{6pt}
\noindent\textbf{Declaration of interests.} The authors report no conflict of interest.

\vspace{6pt}
\noindent\textbf{Data availability statement.} The QuICC framework is openly available at \url{https://github.com/QuICC}. The data that support the findings of this study, and the model configuration used here, are available from the corresponding author upon reasonable request.

\appendix

\section{Summary of the simulations}
\label{A:table}

Table~\ref{T:data} lists every simulation reported in this article: the flux-based and fixed-temperature reduced Rayleigh numbers, the Ekman number, the Nusselt and vertical Reynolds numbers with their time-averaging uncertainties, and the spatial resolution and time step. Cases are grouped by boundary condition, heating profile and Prandtl number, in the order in which they are discussed in the text.

\input{tables/summary_rapids.tex}

\FloatBarrier

\bibliographystyle{jfm}
\bibliography{journal_abbreviations,References}

\end{document}

%% file: tables/summary_rapids.tex
\begin{table}
\scriptsize
\centering
\begin{tabular}{l @{\hskip 0.13in} l @{\hskip 0.13in} l @{\hskip 0.13in} l @{\hskip 0.13in} l @{\hskip 0.13in} l @{\hskip 0.13in} l}
\hline
\hline
\multicolumn{1}{c}{$\Rat_q$} & \multicolumn{1}{c}{$Ek$} & \multicolumn{1}{c}{$\Rat$} & \multicolumn{1}{c}{$Nu$} & \multicolumn{1}{c}{$\widetilde{Re}_z$} & \multicolumn{1}{c}{$N_x \times N_y \times N_Z$} & \multicolumn{1}{c}{$\Delta t$} \\
\hline
\multicolumn{7}{c}{SF-SF, $Q(Z) = \mathcal{E}_c(0.05,Z)$, $Pr=7$}\\
$10^{2}$ & $0$ & $11.8$ & $3.81\pm0.031$ & $0.335\pm0.0030$ & $192\times 192 \times 72$ & $2\times 10^{-2}$ \\
$5\times 10^{2}$ & $0$ & $16.4$ & $13.7\pm0.24$ & $0.832\pm0.0070$ & $288\times 288 \times 96$ & $5\times 10^{-3}$ \\
$10^{3}$ & $0$ & $18.2$ & $24.7\pm0.42$ & $1.19\pm0.010$ & $288\times 288 \times 96$ & $2\times 10^{-3}$ \\
$2\times 10^{3}$ & $0$ & $20.0$ & $45.0\pm0.59$ & $1.68\pm0.011$ & $384\times 384 \times 144$ & $2\times 10^{-3}$ \\
$5\times 10^{3}$ & $0$ & $26.5$ & $85.0\pm6.6$ & $2.77\pm0.047$ & $384\times 384 \times 192$ & $10^{-3}$ \\
$10^{4}$ & $0$ & $36.7$ & $122\pm14$ & $4.10\pm0.066$ & $576\times 576 \times 288$ & $2\times 10^{-4}$ \\
$2\times 10^{4}$ & $0$ & $56.4$ & $160\pm17$ & $6.02\pm0.098$ & $576\times 576 \times 288$ & $2\times 10^{-4}$ \\
$5\times 10^{4}$ & $0$ & $115$ & $196\pm14$ & $9.23\pm0.18$ & $576\times 576 \times 288$ & $2\times 10^{-4}$ \\
$10^{5}$ & $0$ & $182$ & $248\pm15$ & $12.7\pm0.25$ & $768\times 768 \times 384$ & $2\times 10^{-4}$ \\
$2\times 10^{5}$ & $0$ & $260$ & $347\pm26$ & $19.1\pm0.93$ & $1152\times 1152 \times 432$ & $10^{-4}$ \\
\hline
\multicolumn{7}{c}{SF-SF, $Q(Z) = \mathcal{E}_c(0.05,Z)$, $Pr=1$}\\
$10^{2}$ & $0$ & $12.1$ & $3.72\pm0.058$ & $2.54\pm0.028$ & $192\times 192 \times 72$ & $5\times 10^{-3}$ \\
$5\times 10^{2}$ & $0$ & $18.4$ & $12.3\pm0.70$ & $7.23\pm0.15$ & $384\times 384 \times 96$ & $5\times 10^{-4}$ \\
$10^{3}$ & $0$ & $23.7$ & $19.0\pm1.3$ & $10.7\pm0.28$ & $384\times 384 \times 144$ & $2\times 10^{-4}$ \\
$2\times 10^{3}$ & $0$ & $31.9$ & $28.2\pm2.2$ & $15.1\pm0.45$ & $384\times 384 \times 144$ & $2\times 10^{-4}$ \\
$5\times 10^{3}$ & $0$ & $45.7$ & $49.2\pm4.9$ & $23.8\pm0.88$ & $576\times 576 \times 384$ & $2\times 10^{-4}$ \\
$10^{4}$ & $0$ & $59.2$ & $76.0\pm7.5$ & $33.3\pm1.5$ & $768\times 768 \times 576$ & $10^{-4}$ \\
$2\times 10^{4}$ & $0$ & $73.8$ & $122\pm13$ & $47.9\pm2.3$ & $768\times 768 \times 768$ & $5\times 10^{-5}$ \\
$5\times 10^{4}$ & $0$ & $102$ & $221\pm29$ & $74.3\pm5.7$ & $768\times 768 \times 1152$ & $5\times 10^{-5}$ \\
\hline
\multicolumn{7}{c}{SF-SF, $Q(Z) = 1$, $Pr=1$}\\
$10^{2}$ & $0$ & $15.0$ & $3.33\pm0.048$ & $2.61\pm0.030$ & $192\times 192 \times 96$ & $2\times 10^{-3}$ \\
$5\times 10^{2}$ & $0$ & $25.5$ & $9.79\pm0.40$ & $7.43\pm0.16$ & $384\times 384 \times 96$ & $5\times 10^{-4}$ \\
$10^{3}$ & $0$ & $33.7$ & $14.8\pm0.76$ & $11.0\pm0.28$ & $384\times 384 \times 144$ & $5\times 10^{-4}$ \\
$2\times 10^{3}$ & $0$ & $45.7$ & $21.9\pm1.3$ & $15.5\pm0.47$ & $576\times 576 \times 192$ & $2\times 10^{-4}$ \\
$5\times 10^{3}$ & $0$ & $65.5$ & $38.2\pm3.0$ & $24.3\pm0.84$ & $576\times 576 \times 288$ & $10^{-4}$ \\
$10^{4}$ & $0$ & $86.5$ & $57.8\pm4.8$ & $35.3\pm1.9$ & $576\times 576 \times 384$ & $10^{-4}$ \\
$2\times 10^{4}$ & $0$ & $111$ & $90.1\pm7.2$ & $50.3\pm3.9$ & $576\times 576 \times 288$ & $5\times 10^{-5}$ \\
$5\times 10^{4}$ & $0$ & $152$ & $164\pm13$ & $83.5\pm8.0$ & $576\times 576 \times 576$ & $3\times 10^{-5}$ \\
\hline
\multicolumn{7}{c}{NS-SF, $Q(Z) = \mathcal{E}_c(0.05,Z)$, $Pr=7$: matched to \cite{gH24}, fixed $Ra_q = 10^{12}$}\\
$3.97\times 10^{3}$ & $5\times 10^{-7}$ & $17.5$ & $103\pm2.4$ & $2.38\pm0.018$ & $384\times 384 \times 192$ & $5\times 10^{-3}$ \\
$6.21\times 10^{3}$ & $7\times 10^{-7}$ & $19.8$ & $142\pm6.1$ & $3.01\pm0.033$ & $384\times 384 \times 192$ & $2\times 10^{-3}$ \\
$10^{4}$ & $10^{-6}$ & $24.5$ & $185\pm24$ & $4.00\pm0.053$ & $384\times 384 \times 384$ & $2\times 10^{-3}$ \\
$1.72\times 10^{4}$ & $1.5\times 10^{-6}$ & $30.6$ & $253\pm42$ & $5.37\pm0.074$ & $384\times 384 \times 576$ & $10^{-3}$ \\
\hline
\multicolumn{7}{c}{NS-SF, $Q(Z) = \mathcal{E}_c(0.05,Z)$, $Pr=7$: broader scan at fixed $Ek = 10^{-7}$}\\
$10^{2}$ & $10^{-7}$ & $9.96$ & $4.53\pm0.012$ & $0.331\pm0.0012$ & $192\times 192 \times 192$ & $2\times 10^{-2}$ \\
$5\times 10^{2}$ & $10^{-7}$ & $12.9$ & $17.5\pm0.22$ & $0.830\pm0.0066$ & $192\times 192 \times 192$ & $2\times 10^{-2}$ \\
$10^{3}$ & $10^{-7}$ & $14.1$ & $31.9\pm0.52$ & $1.19\pm0.0095$ & $192\times 192 \times 192$ & $10^{-2}$ \\
$2\times 10^{3}$ & $10^{-7}$ & $15.4$ & $58.4\pm1.1$ & $1.68\pm0.016$ & $288\times 288 \times 192$ & $5\times 10^{-3}$ \\
$5\times 10^{3}$ & $10^{-7}$ & $17.5$ & $128\pm2.8$ & $2.63\pm0.026$ & $288\times 288 \times 192$ & $5\times 10^{-3}$ \\
$10^{4}$ & $10^{-7}$ & $20.5$ & $219\pm9.6$ & $3.76\pm0.044$ & $384\times 384 \times 192$ & $2\times 10^{-3}$ \\
$2\times 10^{4}$ & $10^{-7}$ & $37.8$ & $238\pm39$ & $5.85\pm0.083$ & $384\times 384 \times 576$ & $10^{-3}$ \\
$5\times 10^{4}$ & $10^{-7}$ & $84.5$ & $266\pm17$ & $8.95\pm0.12$ & $384\times 384 \times 768$ & $5\times 10^{-4}$ \\
$10^{5}$ & $10^{-7}$ & $147$ & $307\pm18$ & $12.2\pm0.17$ & $576\times 576 \times 576$ & $5\times 10^{-4}$ \\
$2\times 10^{5}$ & $10^{-7}$ & $236$ & $381\pm23$ & $16.9\pm0.33$ & $576\times 576 \times 768$ & $3\times 10^{-4}$ \\
\hline
\hline
\end{tabular}
\caption{Summary of the simulations. The SF-SF cases are computed with $\ep = 0$, i.e.\ on the $Ek = 0$ manifold. For the NS-SF cases the two Rayleigh numbers are related by $\Rat_q = Ek^{4/3} Ra_q$; within each block the quantity held fixed is exact and the other is quoted to three significant figures. The uncertainties are the standard deviations of the time series. The Uniform case at $\Rat_q = 5\times10^{4}$ was also run at $576\times576\times288$ with $\Delta t = 5\times10^{-5}$ for $12$ time units, giving $\Rat = 159$, $Nu = 157\pm13$ and $\widetilde{Re}_z = 75.7\pm5.5$, within one standard deviation of the tabulated values. Our normalisation of $Nu$ differs from that of \cite{gH24} by a factor of $2.22$; values plotted in figure~\dnsfig\ are renormalised to match.}
\label{T:data}
\end{table}